# Incentives Against Power Grabs

or How to Engineer the Revolution in a Pooled Proof of Stake System

Aggelos Kiayias*   Elias Koutsoupias†   Aikaterini-Panagiota Stouka ‡

Tuesday 16th November, 2021

**Abstract**

Proof-of-Stake (PoS) blockchain systems, especially those that allow stakeholders to organize themselves in "stake-pools", have emerged as a compelling paradigm for the deployment of large scale distributed ledgers. A stake-pool operates a node that engages in the PoS protocol and potentially represents a large number of smaller stakeholders. While such pooled PoS operation is attractive from various angles, it also exhibits a significant shortcoming that, so far and to the best of our knowledge, has not been sufficiently understood or investigated. Pooled PoS operation, to be effective and not lead to sub-optimal dictatorial or cartel-like configurations, should enable the stakeholders to revoke and re-delegate their stake in a way that is aligned with their incentives. However, given that stake-pool operators are exactly those entities who determine what transactions are to be recorded in the ledger, they are quite likely to form a cartel and censor any transaction they want, such as those that attempt to adjust the current stake-pool lineup. In this way, a power grab takes place, where the stake-pool cartel perpetuates its control over the PoS system.

We first model and observe formally the emergence of the above problem in pooled PoS systems. Then, we describe an anti-censorship mechanism that takes advantage of the underlying cryptographic functions of the ledger and the nature of peer-to-peer networks to diffuse information without suppression. Specifically, the mechanism exploits digital signatures which can endorse messages that are censored and bind them to delegation transactions, as well as the ability of the underlying ledger to produce unpredictable randomness which can be used to randomly "audit" the suitability of a stake-pool by forcibly expiring it. Upon such expiration, members can evaluate their membership taking into account what information is available, including information potentially not recorded in the ledger but diffused in the peer-to-peer network. We provide a thorough game-theoretic analysis of this mechanism discovering various types of Nash equilibria which demonstrate that the "revolution", i.e., the strategic decision of pool members to withdraw support from a censoring cartel as well as the pool operators to step down, can be incentivized, under suitable and plausible conditions in the utility functions of the involved participants.

# 1 Introduction

Decentralisation is one of the fundamental characteristics of the blockchain and distributed ledger space — arguably it might be the most essential one. For this reason, it is rather striking

---

*University of Edinburgh & IOHK, `akiayias@inf.ed.ac.uk`
†University of Oxford, `elias.koutsoupias@cs.ox.ac.uk`
‡University of Edinburgh & IOHK, `A.Stouka@ed.ac.uk`

that many deployed systems exhibit a rather small level of decentralisation.[1]

While surprising at first, this current state of affairs is less so when put under scrutiny. Recent works such as [6, 17, 2, 18] showed that various classes of mechanisms that organize a set of rational stakeholders around a common goal can easily lead to centralized dictatorial solutions or a low degree of centralization.

In terms of their design approach, there are two major methods in building large scale distributed ledgers. The first one suggested by Bitcoin [20] and followed by a large number of other projects relies on the concept of proof-of-work (PoW) [11, 4] to coordinate participation in ledger maintenance. The main criticism against Bitcoin is its high-energy expenditure. To address the problem of energy expenditure, a second method has been proposed early on that employs proof-of-stake (PoS), see e.g., [19, 16, 10, 7]; in a PoS system participants engage in the protocol based on the amount of stake recorded in the ledger. While energy expenditure is minimal, the problem of incentivising decentralized operation in PoS systems remains an active area of research and it is also the focus of the present work. The flipside of the above line of reasoning that advocates for decentralization is the realization that some degree of centralized operation can be beneficial in a distributed ledger. The main rationale is that offering a minimum level of quality of service (QoS), requires effort and a highly decentralized system, be it PoW-based or PoS-based, may not attract sufficiently equipped or knowledgeable participants to deliver the level of QoS that would make the system useful to the participants. For this reason the vast majority of PoW and PoS systems have mechanisms, either built-in or as overlays, that enable parties to aggregate themselves into groups and either work together forming mining pools, or vote with their stake to elect some well equipped participants to perform the protocol on their behalf.

In the PoS setting, stakeholders of the ledger can organize themselves in block producing entities (hereinafter referred to as "stake-pools") by simply issuing a certificate recording the details of their pool and then record it the ledger. Ethereum-2.0's "validators" is an example of such an approach.[2]

Given that not all nodes are willing or able to setup a block producing entity, many PoS systems allow for "delegating" one's stake to an already registered pool so that the pool can represent them in block production; this gives rise to what we call a *Pooled PoS system*. Such a system simplifies coordination between stakeholders, who otherwise would have to organize themselves via an off-chain protocol. The act of delegation itself can be also very simple, merely adding a certificate to the blockchain with the details of the stake-pool receiving the stake; once the delegation certificate is deposited, there is no need for stakeholders to engage in the protocol and they may be offline except for periodically checking the system to obtain their rewards. Most major PoS systems adopt this delegated approach; we mention a few examples that have been widely discussed in public forums and include Bitshares' "witnesses" [5], Tezos' "bakers" [22], EOS's "block-producers" [12], Qtum's "offline staking" [21], and Cardano's "stake-pools" [15]. Despite the advantages of delegation in the PoS setting, one cannot fail to observe a major shortcoming they possess. The process of vetting stake-pool proposals will inevitably lead to delegation revocations and subsequent reassignment of delegate stake to alternative and more competitive stake-pool proposals. It is in this way that the system is expected to evolve over time and maintain an approximately efficient operation continuously. But if the current line-up of stake-pools is the one responsible for recording such delegated stake reassignments and new pool registrations, *what is to prevent stake-pools to form a cartel and censor all attempts to change the block-producing entities?* After all, it is the stake-pool operators themselves who decide what transactions are to be recorded in the blocks that comprise the distributed ledger, and thus they would have no incentive to add a transaction that removes them from power or allows a competitive pool to be registered. This seemingly unsolvable conundrum is the focus of the present paper.

---

[1]For instance, refer to https://www.bitcoinera.app/arewedecentralizedyet/ for a comparison between popular cryptocurrency systems.

[2]Please refer to https://github.com/ethereum/eth2.0-specs for more details.

In our analysis we assume that the players are *rational*, which means that they act in a way that maximizes their *utility* that is a quantity they find important such as their rewards. The actions that they choose are called *strategies*. A *joint strategy* or a *strategy profile* is a vector with the strategies of all the players. A strategy profile is a *Nash equilibrium* when the following holds for every player $i$: if all the players except $i$ follow the strategy that is indicated by the strategy profile, player $i$ cannot increase its utility by choosing a strategy different from the indicated.

*Our Results:* We initiate the game-theoretic study of censoring transactions in pooled PoS systems, with a particular focus on transactions that modify the delegation of stake or register new pools. In more details, our results are as follows.

(I) We formalize the problem of censorship in pooled PoS systems and introduce the concept of (strong) incentive-consistency for transactions. We observe that for strongly incentive-consistent transactions, liveness holds even in a mixed Byzantine/Rational setting (without any altruistic participants, cf. [1]). On the other hand, we show that there are classes of transactions that are not incentive consistent: such transactions relate to stake re-delegation or pool registration, and there is a Nash equilibrium where they are censored indefinitely by a cartel of stake-pools which is in power.

(II) We put forth an anti-censorship mechanism for such transactions that is motivated by the above results. Our mechanism combines two core ideas. First, a randomized pool dissolving event that happens regularly in the system's operation in an unpredictable fashion, enforced by the ledger rules, capitalizing on the ability of the underlying blockchain to draw unpredictable randomness. Second, the cryptographic compounding of stake delegation transactions which can be strongly incentive consistent with transactions that are non-incentive consistent. This compounding effect offers an opportunity to "break" a stake-pool cartel by challenging its members with the question: do they prefer to leave some stake undelegated (acting as "censors") or include a non-incentive compatible transaction to the ledger? At the same time, it offers the opportunity to the recently dissolved pool members, to either "rebel" against the cartel (and compound a new pool registration transaction that is circulating in the peer-to-peer network to their stake delegation transaction by cryptographically signing it) or "capitulate" and re-delegate their stake to the status-quo operators.

(III) Does the above mechanism actually work? We analyze our anti-censorship mechanism from a game-theoretic perspective. First we provide a single round analysis which reveals the equilibria of the game between the stake-pool operators and their members. It turns out that two distinct equilibria are possible. In the first one, transaction censorship and capitulation takes place, while in the other, assuming the utility of participants aligns properly, facilitates a "revolution", where some pool members choose to rebel and force the stake-pool operator to stop censoring the new stake-pool registration transactions. We subsequently examine how the basic game unfolds over a number of rounds as well as in the setting of adaptive strategies in two or more rounds, where decision-making in one round is adaptive to past game actions. In this setting we show a richer set of equilibria where more players rebel in a round if in a previous round a player has rebelled but was censored. The above results in the multi-round setting are obtained when a single pool is pitted against its pool members. Applying this in practice, will require the multi-round execution to take place in long intervals since in each round a stake-pool is selected at random, thus two rounds pitting the same pool with its members may take an expected $O(k)$ rounds to occur where $k$ is the total number of pools that run the PoS protocol. Motivated by this, we remove this inefficiency, by studying the multi-round setting in the scenario where different pools are assigned to each round.

We remark that the central game-theoretic problem faced by pool members that prefer the formation of a new pool is how to coordinate. Coordination of players to select an appropriate equilibrium is a vast topic in game theory. More relevant to our approach is the investigation of effects of "cheap talk" (see for example [9, 3, 14, 13]) in coordination games [8, 23].

## 2  PoS Systems and Stake Pools

A PoS system is a system supporting a transaction ledger that relies on a PoS mechanism for consensus. Specifically, player $P_i$ has some recorded stake $s_i$ in the system that is associated with a public-key $\mathsf{pk}_i$ and a corresponding secret-key $\mathsf{sk}_i$. In PoS system, at any given time in the protocol execution, parties can determine whether they have to participate in the protocol by using their secret-key $\mathsf{sk}_i$.

The information recorded in the ledger is described by a language $\mathscr{L}$. Players issue transactions $\mathsf{tx}$ that interact with the state of the ledger $x \in \mathscr{L}$ in some way and following a publicly known transition function $\delta$ result in an updated state $x'$, i.e., $\delta(x, \mathsf{tx}) = x'$.

Given that a PoS system is supposed to continuously process transactions, at any given time, in the view of a player $P_i$ there is a ledger $x_i$ that is considered "settled", as well as a sequence of transactions $\mathsf{tx}_1, \ldots, \mathsf{tx}_q$ for which it holds that they extend the settled state in some way $x_i \xrightarrow{\delta, \mathsf{tx}_1} x_{i,1} \ldots \xrightarrow{\delta, \mathsf{tx}_q} x_{i,q}$, and are pending. We say that $x \preceq x'$ when there is a valid sequence of transactions that under $\delta$ transitions $x$ to $x'$. Not all transactions are admissible for a given ledger state; when a transaction is inadmissible, the state of the ledger remains identical, i.e., $\delta(x, \mathsf{tx}) = x$.

A PoS system provides the following two properties.

**Definition 1. Consistency.** *At any two times $t \leq t'$ during the course of the system's operation and any two players $P, P'$ with ledger states $x, x'$ respectively at times $t, t'$, it holds that $x \preceq x'_q$, where $x' \xrightarrow{\delta, \mathsf{tx}_1 \ldots \mathsf{tx}_q} x'_q$ and $\mathsf{tx}_1, \ldots, \mathsf{tx}_q$ are the pending transactions in the view of $P'$ at time $t'$, i.e., ledger state $x'_q$ is a valid extension of ledger state $x$.*
**Liveness.** *If a transaction $\mathsf{tx}$ is given as input to all honest players at a certain time $t$, then at time $t + u$, any honest player $P$, is at a state $x$ for which it holds that there exists some $x_1 \preceq x_2 \preceq x$ and $x_1 \xrightarrow{\delta, \mathsf{tx}} x_2$, i.e., transaction $\mathsf{tx}$ has been processed and its results are incorporated in ledger state $x$. We call $u$ the Byzantine liveness parameter of the PoS system.*

Aside of the above two properties, the exact syntax of transactions, their generation mechanism as well as the specifics of the transition function $\delta$ are not important in the present exposition and we will not constrain them further except that we will make the assumption that each player can use their secret key $\mathsf{sk}_i$ to issue a special transaction denoted by $\{M\}_i$ for some string $M \in \{0,1\}^*$ that results in the public state updated with in such a way that the message $M$ is recorded in the ledger. Note that writing the message $M$ in the ledger will incur a cost (typically proportional to the length of the message $|M|$) and will thus affect the stake of the player (albeit in a certain minor way).

Being online, in general, incurs a cost and as a result it cannot be expected that players will be online all the time. For this reason it has been suggested and deployed in a few different systems (see e.g., [6] and references therein, notably EOS [12], Tezos [22], and Cardano [15]), that the ledger is used to record "stake pool" registration messages, that record a message as follows:

$$G_{i,\mathsf{pname}} = \{\text{New pool with parameters } \mathsf{pname}, \mathsf{pk}\}_i$$

that associates a public-key $\mathsf{pk}$ with a name $\mathsf{pname}$ (possibly other pool related data are added to this record — we omit them in this simplified exposition). This registration mechanism, enables parties to delegate their stake (or part of it) to pools they desire by issuing messages of the form

$$D_{i, a \to \mathsf{pname}} = \{\text{I delegate } a \text{ of my stake to pool } \mathsf{pname} \| \mathsf{nonce}\}_i$$

where nonce is a unique identifier for the stake-pool delegation message. We note that the (pk,sk) pair employed by a stake-pool would be of the same form as the keys used for regular accounts in the system. A delegation message can be subsequently revoked by issuing a message of the form

$$R_{i,\text{nonce}} = \{\text{I revoke my delegation corresponding to nonce}\}_i$$

The above modifications imply a *pooled* version of the proof-of-stake protocol that operates as follows. First of all, each player that has delegated their stake can be offline. A stake pool operator will engage in the protocol using their stake delegated to their pool as opposed to the stake they actually possess. The protocol now can operate in a number of different ways: (i) in the strict pool setting, only pools with delegated stake are allowed to engage in the protocol, (ii) in the mixed setting, players and stake pools are allowed to operate concurrently.

In more details, each player (stakeholder or stake pool operator) parses the state $x \in \mathscr{L}$ for messages of the form $D_{i,a \to \text{pname}}$ and organizes them in a stake-pool lookup table. The table is continuously updated as the state evolves. At any time, when a protocol message arrives that is associated with a stake pool operator or player it is verified with respect to the stake-pool lookup table and the stakeholder distribution. Stakeholders whose stake is delegated are not permitted to participate (up until the moment the revoke their delegation message).

## 3 The Censorship Problem

The censorship problem in a pooled proof-of-stake system refers to the attack scenario where pool operators filter incoming transactions and in this way control the evolution of the ledger both in terms of what is admissible into the ledger state but also in terms of changing the composition of the running pools. This is a serious consideration for the following reason: PoS protocols are typically shown to satisfy consistency and liveness as long as a Byzantine adversary controls a certain amount of stake below a certain threshold (typically this threshold is $1/2 - \epsilon$ or $1/3 - \epsilon$ where $\epsilon > 0$ is a small constant depending on the network characteristics). In a pooled PoS system the responsibility is transfered to the pool operators and hence it can be significantly easier to create a coalition that deviates from the protocol in a way that consistency, and most importantly (for the issue of censorship) liveness is violated.

For regular transactions the above problem is typically mitigated via a simple incentive mechanism that somehow incorporates transaction fees into rewards given to the operators. In order to make this explicit consider the following setting. Each rational player $P_i$ in the system is associated with a utility function that it tries to maximize. The utility function $u_i : \mathscr{L} \to \mathbb{R}$ maps ledger states to a particular real value. We remark that there are typically more factors that influence the utility of a player but, for simplicity, and without loss of generality for our exposition we will just compound all to the ledger state.

When a transaction tx is considered by a PoS player who maintains the ledger and is about to issue a block, a two option strategy is at play, either *include* tx or *censor* tx. If a player has a certain ledger $x$ for which it holds that $u_i(x) < u_i(x')$ where $x \xrightarrow{\delta, \text{tx}} x'$ then the transaction could be possibly included. We remark that censoring is merely postponing a transaction and a PoS player may want to reconsider it at some later time. It is easy to see that the above condition on the utility of a player is not sufficient to establish a formal anti-censorship result. The reason is that while momentarily a transaction may be beneficial to the player (e.g., because it offers some fee), the transaction (i) may not be beneficial in the long run and as a result a non-myopic player (a player that tries to maximize its utility in the long term) will decide to censor it, (ii) it may be more beneficial to postpone it in favor of other transactions that increase the utility even higher. We introduce the following concept to classify transactions from the players' perspective.

**Definition 2.** *A transaction* tx *is* strongly incentive-consistent *(resp. incentive-consistent) if it holds that at any time $t$ of an execution of a PoS system, it holds that for all players $P_i$, and all $x_1, x_2, x'$ such that $x \xrightarrow{\delta,...\not\ni\text{tx}} x_1 \xrightarrow{\delta,...\text{tx}...} x'$ and $x \xrightarrow{\delta,...\text{tx}...} x_2 \xrightarrow{\delta,...} x'$, we have that $u_i(x_1) < u_i(x_2)$ (resp. $u_i(x) \le u_i(x')$), i.e., including the transaction* tx *sooner than later makes player $P_i$ strictly better off in terms of utility in* any *possible extension of its present settled ledger $x$, (resp. including the transaction* tx *reduces no-one's utility).*

Based on strong incentive-consistency, liveness can be derived as follows.

**Theorem 1.** *Any PoS system, if it is maintained by a population of rational stakeholders, satisfies liveness for strongly incentive-consistent transactions as a Nash equilibrium, even in the presence of a Byzantine adversary below the system's Byzantine threshold.*

*Proof.* Observe that the population of participants is divided in rational and Byzantine ones. The rational participants possibly deviate from the protocol with respect to liveness deciding whether to include or postpone an incoming transaction tx. Let us examine now the strategy to include incentive-consistent transactions as they appear. It is easy to see that due to definition 2 no rational participant gains by deviating and postponing such transaction. It follows that including such transactions is a Nash equilibrium strategy independently of the strategy of the Byzantine participants. □

It may come as a surprise that incentive-consistency by itself is not enough to provide liveness. The reason is that merely assuming the monotonicity of the utility function is not enough to incentivize the inclusion of a transaction; for instance, the utility may be such that it favors the delay of a transaction and it grows more steeply as long as a transaction is not included up to a certain delay window. To make matters worse some transactions may not even be incentive-consistent. In particular, transactions that revoke a stake-pool delegation or register a new pool can be arguably seen to be inherently *not* incentive-consistent. The reason is that being a stake-pool enables the operator to wield a certain power over the system's operation that could make his utility to drop when their pool has some delegated stake removed or a competitive pool is registered. We formulate this as the incumbency property below.

**Incumbency Property.** In a PoS system, there exist stake pool delegation revocation transactions and pool registration transactions that are inherently non-incentive-consistent.

In particular, regarding the revocation transactions, for any ledger state $x$ there exists $a \in (0,1)$, such that $x$ can include $\{(\text{pk}, \text{pname})\}_i, D_{j,a \to \text{pname}}$, for some $\text{pname}, \text{pk} \in \{0,1\}^*$ and $i,j \in \{1,\ldots,n\}$ and in such case, it holds that the utility of the operator $P_i$ of pool pname satisfies $u_i(x) > u_i(x')$ for any ledger state $x'$ such that $x \xrightarrow{\delta,\text{tx}} x'$ and $\text{tx} = R_{j,\text{nonce}}$ where nonce identifies transaction $D_{j,a \to \text{pname}}$.

The above property suggests that transactions that revoke a delegation instruction for a certain amount of stake $a$ will always have one dissenter among the stake pool operators (the operator whose pool is "hurt" by having some of its delegated stake removed). This operator will prefer to postpone that transaction; in fact, they will want to postpone it indefinitely. While the above argument points to the existence of a single dissenter, it is plausible in a real-world setting that a coalition of all pools will be formed which, acting as a cartel, will collectively reject all revocation messages submitted by stakeholders, hence perpetuating their custody over the PoS system in defiance to the will of the stakeholders.

Pool registration transactions are also *not* incentive-consistent; *all* the existing stake pool operators have incentives to postpone these transactions, as their utility will decrease if they include them due to competition. For example, according to the reward sharing mechanism proposed in [6], in an equilibrium the $k$ players with the best characteristics in terms of *cost efficiency* and stake open a pool and choose their *margin* (the fraction of the pool's profit they charge their members for their service), so that their pool is more competitive than any other pool that can be registered. However, if the stake pool operators are the ones who register new pools by including pool registration transactions in their blocks, then they have incentives to postpone all the pool registration transactions. This happens because if they postpone all these transactions, then they can set a very high margin that increases their utility, as they are not afraid of losing members due to competition. This reasoning proves that there is a Nash equilibrium where no operator includes pool registration transactions and the liveness is get hurt even with no Byzantine players.

## 4 Our Anti-Censorship Mechanism

The core idea behind our mechanism is as follows: based on Theorem 1 we have liveness for strongly incentive-consistent transactions. A candidate class of transactions for strong incentive-consistency in a PoS system are delegation transactions $D_{i,a \to \mathsf{pname}}$ that delegate some stake that was undelegated in ledger state $x$. Any delay in processing such a transaction can be assumed to be detrimental to all players due to loss of rewards.[3] With this as a starting point, we will explore how we can capitalize on such transactions to ensure liveness for transactions that are not incentive-consistent. Motivated by the incumbency property, we will focus on transactions of the form $R_{j,\mathsf{nonce}}$ and $G_{j,\mathsf{pname}}$.

In more details, we will take advantage of the ability of the distributed ledger to produce unpredictable randomness in regular intervals (we will call them rounds). In each round the ledger is assumed to produce an unpredictable number $\rho$ out of which, a single random stake pool can be selected uniformly and independently. We will dissolve this stake pool as a form of "randomized audit" and will invite its former members to join it. The members will have to come online and engage in delegation, for otherwise they risk loosing rewards.

In addition to the above randomized audit, our mechanism introduces compounding stake delegation transactions with transactions that are possibly not incentive-consistent for the stake-pool operators, in particular revocation and pool registration transactions. Specifically, we allow delegation transactions to include a registration certificate of a new pool, not necessarily the pool in which the stake is delegated, as well as a delegation revocation transaction which is supposed to be processed before the delegation action takes place (hence, someone can revoke and re-delegate their stake in one transaction).

$$[D_{i,a \to \mathsf{pname}} | R_{i,\mathsf{nonce}'} | G_{j,\mathsf{pname}'}] =$$
$$\{\text{I delegate } a \text{ of my stake to pool } \mathsf{pname} \| \mathsf{nonce}$$
$$\| \text{I revoke my delegation corresponding to } \mathsf{nonce}'$$
$$\| \{\text{New pool with parameters } \mathsf{pname}, \mathsf{pk}\}_j\}_i$$

(1)

It is important to note that although we allow $i = j$ and $\mathsf{pname} = \mathsf{pname}'$, this is not necessary; in fact, the mechanism allows a stake delegation message to carry a pool registration certificate that was issued separately by a different stakeholder. We note that the same mechanism would apply to any other type of transaction that is at risk of censorship, nevertheless, we focus on the above type of transactions since they constitute a most obvious target of censorship.

The main intuition behind the mechanism is the following. *By dissolving a stake-pool and subsequently allowing the stake delegation messages to carry new pool registration messages, an opportunity is created to circumvent the incumbency property.* In particular consider a stake pool operated by player $P_i$, which is dissolved for a randomized audit in ledger state $x$. We observe that processing a delegation revocation transaction over $x$ which removes stake from the pool of $P_i$, does not reduce the utility of $P_i$ (since the pool is dissolved at that point and receives no rewards), and moreover, new delegation transactions can be conditional to the introduction to new pools that create new opportunities for delegation. As a result, there exist possible extensions of the ledger state $x$, where another player $P_j$ introduces a new pool, which may be more profitable for a coalition of members that includes $P_i$.

---

[3]This can be facilitated by e.g., calculating the total rewards available as a linear function of total stake delegated.

Given this possibility, the question faced by pool operator $P_i$ is whether to censor stake delegation transactions that carry $P_j$'s pool registration certificate, in an attempt to prevent the creation of the new more attractive pool, or to accept the transaction and either join the new (or some other) pool as a member or retain its pool and adapt its parameters such as its margin [6]. When the pool operator $P_i$ decides to censor the transaction, it essentially rejects all its members that delegate stake to $P_i$ and try to register $P_j$. As a result, $P_i$'s utility is reduced, but it achieves to prevent the creation of the new pool. At the same time, the members that try to register $P_j$, will fail to delegate their stake and will lose some income, hence this action of rebellion against the old stake-pool leader will come with the possibility of a penalty in terms of their utility.

To summarize the game-theoretic issues, we observe that pool operator $P_i$ has to decide whether to censor all transactions that register a competing pool $P_j$ and the pool members have to decide whether to include such registration with their delegation transaction. This game is at the heart of our considerations in this work. However, the situation is significantly more complicated for the following reasons: First, the utilities of both the pool operator and the pool member depend on many other factors. Second, the game is played repeatedly in a potentially dynamic environment. Third, operators and members of the remaining pools can affect dramatically the outcome. Specifically, it takes a single pool operator to accept the registration of a new pool. Finally, there are multiple equilibria in general, and the equilibrium selection process is greatly affected by the socially acceptable behavior. For example, a reasonable social policy would ask players, when indifferent, to include some new pool registration in their delegation transaction.[4]

While it is beyond the scope of the current exposition to provide explicit reward mechanisms for pools to receive rewards, recall that several recent works have done so already, cf. [6, 17], and can be used to support the above reasoning.

To analyze our mechanism we will consider an abstract game that possibly carries over multiple rounds with players choosing whether they want to comply with the existing PoS system configuration into stake pools or support its evolution as resulting from the introduction of new stake pool registrations. Observe that this is inline with a PoS system operation where transactions are assumed to be communicated in an underlying peer-to-peer network and are thus accessible to all players despite the fact that the players that operate the current set of stake pools may choose to censor them and postpone their inclusion into the ledger state. In our analysis we will consider two simplified utility notations, $u, u'$, that capture respectively the utility of a player under the current stake pool status and the one that is emerging from the PoS system after the new pool registrations are included in the ledger. Our objective is to analyze the conditions on $u, u'$ under which including the transactions (and as a result evolving the system) is an equilibrium, as well as studying the equilibrium strategies themselves and their viability.

## 5   Single Round Game

In this section we consider a simplest case of the game-theoretic situation, in which we focus on a single pool and a single round. The game is played between a pool leader and the members of the pool. The interesting case arises when one of the members or somebody else wants to create a new pool, which is potentially more profitable for the members of the existing pool. Therefore, the members would prefer to abandon the current pool and join the new pool. However, the current pool leader may try to resist the creation of the new pool by blocking its registration. We consider the worst case that every other pool operator also blocks the registration of the new pool, which is a reasonable given that a more competitive pool can only reduce their reward in general.

---

[4] An analogous social advice for block mining is to mine at the deepest branch in the setting of "longest chain wins" blockchain protocols.

The pool leader has two possible strategies: censor and notcensor, that is whether to accept some —arguably all— stake delegation transactions that register the new pool or to reject all of them. On the other hand, each pool member has two strategies: capitulate, in which case it posts a transaction for delegating its stake to the old pool without including the registration of the new pool, or rebel, in which case it adds the registration of the new pool (and possibly changes its delegation preference).

In effect, the pool leader poses the following dilemma to its members: (i) to include the new pool registration in their delegation certificate and risk rejection, which means that their stake will temporarily remain undelegated and get no income or (ii) to not include it and continue getting income from delegation, but lose a higher income in the long run from the new pool. We assume that if the pool leader selects notcensor and the new pool is created, all members will have sufficient time to change their delegation and join the new pool independently of whether they rebel or not.

Based on the above reasoning we define the strategies and the utilities of the players and we also find all the pure Nash equilibria of this game. Recall that a utility function takes as input the joint strategy (a vector with the strategies of all the players) and outputs the quantity that the players want to maximize such as profit.

*Strategies of the players:* The set of strategies for the pool leader is {censor, notcensor} and the set of strategies for the pool members is {capitulate, rebel}. The strategy of the pool leader will be denoted by $S_P$ and the strategy of a pool member $i$ will be denoted by $S_i$.

*Utilities:* Suppose that $\alpha$ is the stake of the pool members that select capitulate. The utility of the pool leader is $u_\alpha^P$, when it plays censor, and $u'^P$ when it plays notcensor and at least one member chooses rebel. The first case captures the setting when the pool leader censors all the stake delegation transactions that register the new pool, so the new pool is not registered, and it loses all the members who chose rebel. The latter case captures the setting when revolution takes place, the new pool registration is installed and the stake-pool line up of the system is modified. The case of playing notcensor when all members choose capitulate is captured by the first case (for $\alpha$ that is equal to the total stake of the pool), as the new pool is not registered and no player is censored. Similarly, let $u_\alpha^i$ be the utility of member $i$ when (i) the pool leader chooses censor (ii) $i$ chooses capitulate, and let $u'_i$ be the utility of pool member $i$ when the pool leader selects notcensor and at least one player chooses rebel. The first case gives the utility that member $i$ obtains by being a member of a pool that has total delegated stake $\alpha$, as the members that chose rebel have been censored. The latter case captures the setting when the revolution takes place, a new pool is registered and a pool member can benefit from the new stake-pool lineup.

Let $n$ be the number of players, $s_i$ the stake of member $i$, $s_P$ the stake of the pool leader and $y = s_1 + ... + s_n + s_P$. Using the above notation we define the utility of the pool leader and the pool members in our game as follows: The utility of the pool leader $U_P$ in a joint strategy $(S_1, S_2, ..., S_n, S_P)$ will be

$$U_P(S_1, S_2, ..., S_n, S_P) = \begin{cases} u_\alpha^P & \text{if } S_P = \text{censor} \\ u'^P & \text{if } S_P = \text{notcensor} \land \exists i \, S_i = \text{rebel} \\ u_y^P & \text{if } S_P = \text{notcensor} \land \forall i \, S_i = \text{capitulate} \end{cases}$$

The utility of the pool member $i$ denoted by $U_i$ in a joint strategy $(S_1, S_2, ..., S_n, S_P)$ will be

$$U_i(S_1, S_2, ..., S_n, S_P) = \begin{cases} 0 & \text{if } S_P = \text{censor} \land S_i = \text{rebel} \\ u_\alpha^i & \text{if } S_P = \text{censor} \land S_i = \text{capitulate} \\ u_i' & \text{if } S_P = \text{notcensor} \land \exists i\ S_i = \text{rebel} \\ u_y^i & \text{if } S_P = \text{notcensor} \land \forall i\ S_i = \text{capitulate} \end{cases}$$

Note that when the pool leader chooses notcensor and all the players capitulate, the outcome is that, independently of the presence of the new pool, pool members will remain with the old pool and hence the utility of a player $i$ will be $u_y^i$ not $u_i'$.

The following theorem characterizes the equilibria of this game. It states that there are two types of equilibria: a unique equlibrium in which censor is successful and every member capitulates, and a class of equilibria in which members with sufficient stake can manage to force the pool leader not to censor.

**Theorem 2.** *Let $F = K_1 \cap K_2$ where $K_1$ is the event that there exists unique $i$ such that $S_i =$ rebel and $K_2$ is the event that $(S_i = \text{rebel}) \Rightarrow (u_i' \geq u_y^i)$.*
*Let $J$ be the event that there exist more than one players with rebel. Assuming that (i) $\forall i$ it holds $u_x^i > 0$ for $x \geq s_i + s_P$ and (ii) there exists a player $i$ for whom it holds $u_i' > u_y^i$ all the equilibria of our game are the following:*

- $A = (S_1, ..., S_n, \text{notcensor})$ *such that*

$$\left(u'^P \geq u_\alpha^P \text{ where } \alpha = \sum_{j: S_j = \text{capitulate}} s_j\right) \cap (J \cup F)$$

- $B = (S_1, ..., S_n, \text{censor})$ *such that* $\left(\forall i\ S_i = \text{capitulate}\right)$.

*Proof.* First we will prove that joint strategy $A$ is an equilibrium. If the pool leader chooses censor then their utility will become $u_\alpha^P$ where $\alpha = \sum_{j:S_j=\text{capitulate}} s_j$ which is at most their current utility that is $u'^P$ given that there exists at least one player with rebel.

Regarding the other players: (i) if $F$ holds then if the player $i$ who has chosen rebel changes their strategy to capitulate then their utility will become $u_y^i$ which is at most his current utility which is $u_i'$. If a player $j$ with $S_j =$ capitulate chooses rebel then their utility will not change because already one player has chosen rebel and they have not been censored. (ii) If $J$ holds: then if a player $j$ with $S_j =$ capitulate chooses rebel strategy then their utility will not change. The same happens with a player who has chosen rebel.

Now we will prove by contradiction that there is no other equilibrium than $A$ where the pool leader has chosen notcensor. Let us assume that there is an equilibrium where the pool leader has chosen notcensor but the condition ($u'^P \geq u_\alpha^P$ where $\alpha = \sum_{j:S_j=\text{capitulate}} s_j$) does not hold. Then the pool leader can increase their utility from $u'^P$ to $u_\alpha^P$ by changing his strategy to censor. Note that the utility of the pool leader will be $u'^P$ because we can prove by contradiction that there is no equilibrium where all players have chosen capitulate and the pool leader notcensor. Specifically if there was such an equilibrium then the player for whom it holds $u_i' > u_y^i$ could increase their utility by choosing rebel.

Let us assume that that there is an equilibrium where the pool leader has chosen notcensor but the condition ($F \cup J$) does not hold. This means that either (a) there exist a unique player with rebel strategy but for this player it holds $u_i' < u_y^i$ or (b) all players have chosen capitulate. In case (a) the player with rebel can increase their current utility $u_i'$ to $u_y^i$ if they choose capitulate and in case (b) the player for whom it holds $u_i' > u_y^i$ can increase their utility by choosing rebel.

Now we will prove that joint strategy $B$ is an equilibrium. If the pool leader changes strategy and chooses notcensor then their utility will remain the same because all the players have chosen capitulate. If a player changes their strategy and chooses rebel then their utility will become zero that is smaller than their current utility $u_y^i$.

Now we will prove by contradiction that there is no equilibrium where the pool leader has chosen censor and it is different from $B$. If there was at least one player with rebel then they could increase their utility from zero to $u_x^i$, where $x \geq s_i + s_P$ by choosing capitulate. □

## 6 Extension to Multiple Rounds

In this section we extend the game to multiple rounds. The strategy of each member is to choose capitulate or rebel for each round and the strategy of the pool leader to choose notcensor or censor for each round. We will assume that when the pool leader chooses notcensor or a member chooses rebel in one round then in the following rounds they are restricted to do the same. In the following section we remove this assumption. The utility of a player $i$ for a game with $m$ rounds is $u_i = u_{i,1} + ... + u_{i,m}$ where $u_{i,j}$ the utility of player $i$ in round $j$. If the pool leader chooses notcensor in a round and there exists a player with rebel then in all the following rounds the utility of the members will be $u_i'$ and the utility of the pool leader $u'^P$. Otherwise the utilities of each player during each round are defined as in the one shot game. The players decide in the beginning their strategies for all the rounds. In the next section we will examine also adaptive strategies.

**Definition 3.** *Let $S_{i,j}$ be the strategy of player $i = 1,...,n$ for round $j = 1,...,m$. Let $\vec{S}_i^r$ be the vector with the strategies of player $i$ until round $r$ (including $r$) and $\vec{S}_i^{-r}$ be the vector with the strategies of player $i$ after round $r$. Let $\vec{S}^r = (\vec{S}_P^r, \vec{S}_1^r, ..., \vec{S}_n^r)$ be the vector with the joint strategy until round $r$ and $\vec{S}^{-r} = (\vec{S}_P^{-r}, \vec{S}_1^{-r}, ..., \vec{S}_n^{-r})$ the vector with the joint strategy after round $r$.*

**Definition 4.** *A subgame $r$ will be a game as described above beginning from round $r+1$ and ending after round $m$.*

**Definition 5.** *A joint strategy $\vec{S}^m$ will be a Subgame Perfect Nash Equilibrium if for every subgame $r$ $\vec{S}^{-r}$ is a Nash Equilibrium.*

The following theorem describes a class of joint strategies that are Subgame Perfect Nash Equilibria and extensions of the single-round equilibria. Specifically the class contains all equilibria in which during an initial phase everyone stays in the old pool (i.e., the pool leader chooses censor and all the members capitulate) and at some point the "revolution" succeeds and everyone joins the new pool (i.e., the pool leader chooses notcensor while some members rebel). This class of joint strategies that are Subgame Perfect Nash Equilibria include also the extreme cases where the initial phase is empty or covers all rounds.

**Theorem 3.** *Assume that every member has strictly positive utility when it stays in the old pool, i.e. $u_x^i > 0$, where $x \geq s_i + s_P$, and that the pool leader prefers to keep its pool intact than dissolve it, i.e., $u'^P \leq u_y^P$.*

*The following joint strategies for the game of $m$ rounds are Subgame Perfect Nash Equilibria for every $l \in \{1,...,m\}$: The pool leader selects $S_{P,j} =$ censor, at every time step $j \leq l$, and $S_{P,j} =$ notcensor for $j > l$. All pool members select $S_{i,j} =$ capitulate for $j \leq l$. At time $j = l+1$, there exist either two or more members that select rebel or a unique player $i$ that selects rebel; in the latter case, player $i$ must prefer the new pool from the current situation, i.e., $u_i' \geq u_y^i$, where $y$ is the total stake of the old pool. Furthermore, the total stake of rebel members at time $l+1$ must make the current pool unattractive to the pool leader, i.e., $u'^P \geq u_\alpha^P$ where $\alpha = \sum_{i: S_{i,l+1} = \text{capitulate}} s_i$.*

*Proof.* We will prove that these joint strategies are Subgame Perfect Nash Equilibria. In subgame $m-1$ (which includes only round $m$) the joint strategies $\vec{S}^{-(m-1)}$ are Nash equilibria as we have proved in the previous theorem. Now we will take an arbitrary subgame $m-i$ and we will prove that $\vec{S}^{-(m-i)}$ as described above is a Nash equilibrium of this subgame. $\vec{S}^{-(m-i)}$ can be:

1. in all rounds pool leader has chosen notcensor and all the conditions described in the theorem for that case hold.

2. in some rounds in the beginning pool leader has chosen censor and all the members capitulate and from one point and after the pool leader choose notcensor and the relevant conditions hold.

3. in all rounds pool leader has chosen censor and all the members capitulate.

**In the first case**:

- if the pool leader changes their strategy and chooses censor initially in some rounds (or in all rounds) then (i) their utility for these rounds will not increase because it holds $u'^P \geq u_a^P$ where $a = \sum_{i:s_{i,j}=\text{capitulate}} s_i$ and (ii) their utility in the following rounds will not change because they have chosen notcensor and there exists at least one member with rebel which means that their utility is $u'_P$ regardless what happens in the previous rounds.

- if a pool member chooses rebel in an earlier round then this will not change their utility because already in each round there exist a player with rebel and the pool leader has chosen notcensor. If a pool member $i$ chooses rebel later or not at all, then if there exists other member with rebel in these rounds then their utility will not change. If there is no other member with rebel the utility of $i$ will not increase because $u'_i \geq u_y^i$. Also their utility in the following rounds will be the same.

**In the second case:**

- if the pool leader chooses notcensor earlier then their utility will not change because all the members have chosen capitulate. If the pool leader chooses censor later then their utility will not increase because of the same arguments as above. if a member $i$ choose rebel earlier then their utility will not increase given that the pool leader has chosen censor and it holds $u_y^i \geq 0$.

- if a member chooses rebel later then if there exists other player with rebel then their utility will not change. If there are not any other players with rebel then their utility in these rounds will not increase because $u'_i \geq u_y^i$, and their utility in the next rounds will be the same (because there exist a player with rebel and the pool leader has chosen notcensor).

**In the third case**:

- if the pool leader chooses notcensor then their utility will not change because there is no member with rebel.

- if a member choose rebel then their utility will not increase because $u_y^i \geq 0$.

□

# 7 Adaptive Strategies

In the previous sections we considered non-adaptive strategies and now we turn our attention to a more interesting and realistic setting in which players adapt their strategies based on the situation of previous rounds. The motivation for adaptive strategies is to capture the following scenario: Members of the pool may hesitate to choose rebel because of the threat that the pool leader will select censor and, as a result, they will lose the rewards of the next round. Can the situation be improved by communication? This is a "cheap-talk" situation, in which pre-play communication between players —with no direct affect on the players utilities— does not necessarily lead to a good equilibrium.

We assume that before every round members can freely communicate off-chain and then simultaneously commit to a strategy for the coming round. If members could provide a commitment that they will select the rebel strategy, they could coordinate and "force" the pool leader to select notcensor. But such commitment is impossible to be enforced on chain, as the pool operators would censor them, so members have to rely on "cheap-talk" (with no direct effect on utility) or signals (with some direct effect on utility) to coordinate. A signal that a member can give to others is to select rebel and forfeit the income of the next round. When this happens and the pool leader censors it, the player's choice does not appear on-chain. Still, other players learn about the suppressed strategy from off-chain communication and by checking that it did not appear on-chain. We extend the strategies of the previous section to allow for communication and signaling. Now the strategy of the second round can depend on the strategies selected in the first round.

*Strategies.* The strategy of a member $i$ is $\vec{S}_i = (S_{i,1}, S_{i,2}(S_{1,1}, ..., S_{n,1}, S_{P,1}))$. The strategy of the pool leader is $\vec{S}_P = (S_{P,1}, S_{P,2}(S_{1,1}, ..., S_{n,1}))$. We now define a natural set of strategies $X$ that allow the members to coordinate in the second round by responding to choices of the first round.

- Let $X$ be the strategy of pool members in which they rebel in the second round when they see some member rebel in the first round, i.e., $X$ = rebel if there exists $i$ such that $S_{i,1}$ = rebel, and capitulate otherwise.

- Let $Y$ be the strategy of the pool leader in which they select notcensor in the second round, when they see a rebel in the first round, i.e., $Y$ = notcensor if there exists $i$ such that $S_{i,1}$ = rebel, and censor otherwise.

**Theorem 4.** *Assuming that for all $i$, $u_x^P$ and $u_x^i$ is increasing in $x$ and for all $i$ and $x$, $u_x^i \geq 0$, then the following joint strategies are Nash equilibria:*

- *The strategy of the pool leader is $S_P = (\text{censor}, Y)$, when $u'^P \geq u_{y-l}^P$, where $y$ is the total stake of pool members (including pool leader) and $l = \sum_{i: S_{i,2} = X} s_i$ is the stake of pool members that select strategy $X$.*

- *The strategy of the pool members satisfy $L \cap J \cap F$, where $J$ is the event that $\exists i, j$ such that $S_{i,2} = S_{j,2} = X$, $L$ is the event that there exists unique $i$ such that $S_{i,1} = $ rebel and $F$ the event that $S_{i,1} = \text{rebel} \Rightarrow (u'^P \leq u_{y-s_i}^P) \cap (u_i' \geq 2 u_y^i)$.*

*Proof.* Let $i$ be the member that chooses rebel in the first round. If the pool leader changes their strategy and chooses (notcensor,notcensor) or (notcensor,$Y_1$) where $Y_1$ is arbitrary function that is equal to notcensor for the given strategies of the first round, then the strategy of the pool leader will become $2 \cdot u'^P$ that is not higher than their current utility that is equal to $u_{y-s_i}^P + u'^P$ (because of the first part of $F$ event). If the pool leader chooses (notcensor,censor) or (notcensor,$Y_2$) where $Y_2$ is arbitrary function that is equal to censor for the given strategies of the first round then their utility will become again $2 \cdot u'^P$ that is not higher than their current utility. If the pool leader chooses (censor,censor) or (censor,$Y_3$) where $Y_3$ is arbitrary function that is equal to censor for the given strategies of the first round then their utility will become at most $u_{y-s_i}^P + u_{y-l}^P$ that is not higher than their current utility because of $u'^P \geq u_{y-l}^P$. If the pool leader chooses (censor,$Y_4$) where $Y_4$ is arbitrary function that is equal to notcensor for the given strategies of the first round then their utility will remain the same.

If player $i$ (for whom it holds $S_{i,1}$ = rebel) chooses (rebel,$M$) where $M$ is arbitrary strategy then their utility will remain the same, because there exist at least two members that in the second round have chosen $X$ strategy (event $J$). Note that $X$ in this case is equal to rebel because $S_{i,1}$ = rebel and $Y$ equal to notcensor.

If player $i$ chooses capitulate in the first round (regardless what they choose in the second round) then their utility will become at most $2 \cdot u_y^i$ not higher than $u_i'$ which is their current utility (because of second part of $F$). This holds because if nobody in the first round with rebel exists then the strategy of the pool leader in the second round is censor. We say "at most" because maybe some players have chosen rebel in the second round and $u_x^i$ is increasing in $x$.

Regarding another player $j$ who has chosen capitulate in the first round regardless what this player has chosen in the second round : if they choose (rebel, $M$) where $M$ is arbitrary strategy then their utility in the first round will become zero that is not higher than their current utility $u_{y-s_i}^j$ and their utility in the second round will remain the same. If they choose (capitulate,$M$) where $M$ is arbitrary strategy then their utility will remain the same. □

**Remark 1.** *The equilibria described in Theorem 4 hold even without the assumption that if the pool leader has chosen* notcensor *they will choose* notcensor *again and if a pool member has chosen* rebel *they will choose* rebel *again.*

Next we argue that if there is a signal and enough players who follow the signal the revolution will happen. Intuitively according to the following theorem if there is a player with rebel in the first round and enough players with $X$ in the second round which means that they follow the signal then there is no equilibrium where the pool leader chooses to censor in the second round and prevents the revolution.

**Theorem 5.** *(I) Assuming that $u_x^P$ is increasing in $x$, if $u_P' > u_{y-l}^P$, where $l = \sum_{i:S_{i,2}=X} s_i$, $\exists i$ such that $S_{i,1} =$ rebel then there is no equilibrium where $S_{P,1} =$ censor and $S_{P,2} =$ censor or the pool leader chooses a strategy that ends up to censor in the second round. (II) Assuming that $u_x^i$ is increasing in $x$, if there is a player for whom it holds $u_i' > 2u_y^i$ and there is a player $j$ for whom it holds $S_{j,2} = X$ and $S_{P,2} =$ notcensor or the pool leader chooses a strategy that ends up to notcensor then there is no equilibrium where in the second round the utility of a member $i$ is not $u_i'$.*

*Proof.* For I, let us assume that there exists such an equilibrium. Then the utility of the pool leader will be smaller than $u_y^P + u_{y-l}^P$. If they change their strategy to notcensor in the second round, their utility for the first round will not change and their utility for the second round will become $u'^P$ which is strictly higher.

For II, let us assume that there exists such an equilibrium. Then in the first round there is no player with rebel and in the second round there is no player with rebel or with strategy that ends up to rebel. The utility of player $i$ in this case is at most $2u_y^i$. If this player changes their strategy to rebel in the first round their utility will become $u_i'$ which is strictly higher. □

**Extension to multiple rounds** We can extend theorem 4 so that it captures the above game extended in multiple rounds. If the game lasts $k$ rounds then we can adapt theorem 4 so that (i) the revolution happens in round $j+1$ (ii) for the player who chooses rebel it holds $D_{j,k} = u_i' \geq \frac{k}{k-j} u_y^i$ (instead of $u_i' \geq 2u_y^i$) and (iii) $X, Y$ strategies are adapted as follows:

$X_j =$ rebel if $\exists i$ such that $S_{i,1} = S_{i,2} = ... = S_{i,j} =$ rebel and capitulate otherwise.

$Y_j =$ notcensor if $\exists i$ such that $S_{i,1} = S_{i,2} = ... = S_{i,j} =$ rebel and censor otherwise.

Note that $D_{j,k}$ represents the fact that it is more profitable for a player to rebel for some rounds and get zero rewards compared to capitulating if at the end they manage to transform the system to a new state with a new pool.

**Two round game-two pools** We extend the previous two round game as follows: There are two pools with $n$ players each one active in the first and second round respectively. This captures the scenario where a player of the first pool gives a signal by choosing rebel and in the second round the players of the second pool follow the signal, choose rebel and the revolution takes place. Let $P_1, P_2$ the pool leaders of the pools and $y_1, y_2$ the sums of the stake that belong to all the members and the pool leader of the first and the second pool respectively. The utilities of the players are defined as follows:

- The utility of a player $i$ of the first pool during the first round is defined in the same way as the previous game and in the second round is either $u^i_{y_1}$ or $u'_i$. In the second round the utility of player $i$ becomes $u'_i$ if (i) the pool leader of the second pool chooses notcensor and there is a player of the second pool that chooses rebel (ii) the utility of player $i$ in the first round is $u'_i$.

- The utility of the players of the second pool during the first round is $u^i_{y_2}$. The utility of the players of the second pool during the second round is defined as before except when during the first round $P_1$ chose notcensor and there was a player of the first pool with rebel. In that case the utility will be $u'_i$.

Let $X$ be the strategy of the second pool's pool members where they choose rebel if there exists $i$ such that $S_{i,1}$ = rebel $\cap D$ and capitulate otherwise. Let $Y$ the strategy of the second pool's pool leader where it chooses notcensor if there exists $i$ such that $S_{i,1}$ = rebel $\cap D$ and censor otherwise. Let $D$ be the event that $u'_i \geq 2u^i_{y_1}$.

**Theorem 6.** *Assuming that for all $i$ $u^P_x$ and $u^i_x$ is increasing in $x$ and for all $i$ and $x$ it holds $u^i_x \geq 0$ then the following are Nash equilibria: $S_{P_1}$ = censor, $S_{P_2} = Y$ and $K \cap L \cap J \cap F$ where $K$ is the event that $u'_{P_2} \geq u^{P_2}_{y_2-l}$, where $l = \sum_{i:S_{i,2}=X} s_i$, $J$ is the event that $\exists i, j$ such that $S_{i,2} = S_{j,2} = X$, $L$ is the event that there exists unique $i$ such that $S_{i,1}$ = rebel and $F$ is the event that $(S_{i,1} = \text{rebel}) \Rightarrow ((u'^{P_1} \leq u^{P_1}_{y_1-s_i}) \cap D)$.*

*Proof.* Let $i$ be the player who chose rebel in the first round.

1. If $P_1$ changes their strategy to notcensor then their utility will change from $u^{P_1}_{y_1-s_i} + u'_{P_1}$ to $2u'^{P_1}$ which is not higher because of the first part of $F$.

2. If $i$ changes their strategy to capitulate then their utility will change from $u'_i$ to at most $2u^i_{y_1}$ which is not higher because of $F$.

3. If a player $j$ from the first pool changes his strategy from capitulate to rebel then his utility will change from $u^{P_1}_{y_1-s_i} + u'_j$ to which is not higher.

4. If $P_2$ changes their strategy from $Y$ to censor or to a strategy that ends up to censor then their utility will change from $u^{P_2}_{y_2} + u'^{P_2}$ to at most $u^{P_2}_{y_2} + u^{P_2}_{y_2-\alpha}$ which is not higher. If this pool leader changes their strategy from $Y$ to something that ends up to notcensor then their utility will remain the same.

5. If a player of the second pool who has chosen $X$ changes their strategy their utility will remain the same because there is at least one more player with $X$.

6. If a player of the second pool with a different strategy from $X$ changes their strategy then their utility will remain the same because there exists a player with $X$ and pool leader has chosen $Y$.

□

# Acknowledgment

The first author was partially supported and the third author was supported by H2020 project PRIVILEDGE #780477.